\documentclass{aa}

\usepackage[T1]{fontenc}
\usepackage{pslatex}
\usepackage{amsmath}
\usepackage{amsfonts}
\usepackage{xcolor}
\usepackage{url}
\usepackage{bm}
\usepackage{graphicx}
\usepackage{amssymb}
\usepackage{soul}
\usepackage{booktabs}
\usepackage{array}
\usepackage[utf8]{inputenc}
\inputencoding{latin1}
\inputencoding{utf8}
\usepackage{appendix}

\def\lsim{ \lower .75ex\hbox{$\sim$} \llap{\raise .27ex \hbox{$<$}} }
\def\gsim{ \lower .75ex \hbox{$\sim$} \llap{\raise .27ex \hbox{$>$}} }

\newcommand{\bi}{\begin{itemize}}
\newcommand{\ei}{\end{itemize}}

\usepackage[colorlinks=true,linkcolor=blue,citecolor=blue]{hyperref}

\begin{document}

\title{Multifrequency polarimetry of high-synchrotron peaked blazars\\ probes the shape of their jets} 
\titlerunning{Multifrequency polarimetry of HSP blazars}

\author{
F. Bolis\inst{1,2}
\and E. Sobacchi\inst{2,3}
\and F. Tavecchio\inst{2}
}
\authorrunning{Bolis et al.}

\institute{
DiSAT, Università dell’Insubria, Via Valleggio 11, I-22100 Como, Italy\\
\email{filippo.bolis@inaf.it}
\and 
INAF -- Osservatorio Astronomico di Brera, Via E. Bianchi 46, I-23807 Merate, Italy
\and
Gran Sasso Science Institute, Viale F.~Crispi 7, I-67100 L’Aquila, Italy
}
\date{}

\voffset-0.4in



\abstract{
Multifrequency polarimetry is emerging as a powerful probe of blazar jets, especially with the advent of the Imaging X-ray Polarimetry Explorer (IXPE) space observatory. We studied the polarization of high-synchrotron peaked (HSP) blazars, for which both optical and X-ray emission can be attributed to synchrotron radiation from a population of nonthermal electrons. We adopted an axisymmetric stationary force-free jet model in which the electromagnetic fields are determined by the jet shape. When the jet is nearly parabolic, the X-ray polarization degree is $\Pi_{\rm X}\sim 15-50\%$, and the optical polarization degree is $\Pi_{\rm O}\sim 5-25\%$. The polarization degree is strongly chromatic: $\Pi_{\rm X}/\Pi_{\rm O}\sim 2-9$. This chromaticity is due to the softening of the electron distribution at high energies, and is much stronger than for a uniform magnetic field. The electric vector position angle (EVPA) is aligned with the projection of the jet axis on the plane of the sky. These results compare very well with multifrequency polarimetric observations of HSP blazars. When the jet is instead nearly cylindrical, the polarization degree is large and weakly chromatic (we find $\Pi_{\rm X}\sim 70\%$ and $\Pi_{\rm O}\sim 60\%$, close to the expected values for a uniform magnetic field). The EVPA is perpendicular to the projection of the jet axis on the plane of the sky. A cylindrical geometry is therefore practically ruled out by current observations. The polarization degree and the EVPA may be less sensitive to the specific particle acceleration process (e.g.,~magnetic reconnection or shocks) than previously thought.
}

\keywords{galaxies: jets -- radiation mechanisms: non-thermal -- polarization
}

\maketitle
\boldsymbol{}

\section{Introduction}

Relativistic jets from supermassive black holes in active galactic nuclei (AGNs) shine through the entire electromagnetic spectrum, from the radio up to very high-energy gamma-rays \citep{romero17, Blandford19, Boettcher19}. Nonthermal radiation carries a wealth of information on the dynamics, structure, and composition of AGN jets. Nonthermal radiation is best studied in blazars, a particular class of AGN where the jet is directed nearly along the line of sight, as the emission from the jet is strongly beamed due to the favorable orientation.

Multifrequency polarimetry is emerging as a powerful probe of blazar jets, especially with the advent of the Imaging X-ray Polarimetry Explorer (IXPE) space observatory \citep{liodakis22, digesu22, digesu23, ehlert2023, Marshall2023, middei23b, middei23a, peirson23, Errando2024, Kim2024}. Here we focus on high-synchrotron peaked (HSP) blazars, for which both optical and X-ray emission can be attributed to synchrotron radiation from a population of nonthermal electrons. The observed X-ray polarization degree is $\Pi_{\rm X}\sim 10-20\%$, whereas the optical polarization degree is significantly smaller ($\Pi_{\rm X}/\Pi_{\rm O}\gtrsim 2$). The electric vector position angle (EVPA) seems to align with the projection of the jet axis on the plane of the sky. So far, IXPE has observed blazars in a quiescent state, when the polarization degree and the EVPA do not change significantly during the observing time \citep[see however][]{digesu23}.

Multifrequency polarimetric observations of HSP blazars were interpreted as evidence that nonthermal electrons are accelerated in shocks \citep[e.g.,][]{liodakis22}. Several theoretical studies support this interpretation \citep[for a review, see, e.g.,][]{marscher21, tavecchio21}. However, it is far from evident that the shock acceleration scenario is the only way to interpret observations.

In this paper we study the polarization of the synchrotron radiation from Poynting-dominated jets, where particles are unlikely accelerated by shocks \citep{Sironi15b}. We show that the polarization degree and the EVPA are very sensitive to the jet shape, which determines the global structure of the electromagnetic fields (we neglect the presence of a random component of the fields that changes on small spatial scales). We find that multifrequency polarimetric observations of HSP blazars can be reproduced when the jet is nearly parabolic, whereas a cylindrical geometry is practically ruled out. The softening of the electron distribution at high energies is crucial to explaining the strong chromaticity of the polarization degree. In contrast with previous claims, we argue that current observations can hardly constrain the particle acceleration process.

The paper is organized as follows. In Sect. \ref{sec:structure} we discuss the structure of the electromagnetic fields within the jet. In Sect. \ref{sec:pol} we study the polarization of the synchrotron radiation. In Sect. \ref{sec:disc} we discuss our results and conclude.

\section{Jet structure}
\label{sec:structure}

\begin{figure}
    \centering
    \includegraphics[width=0.75\linewidth]{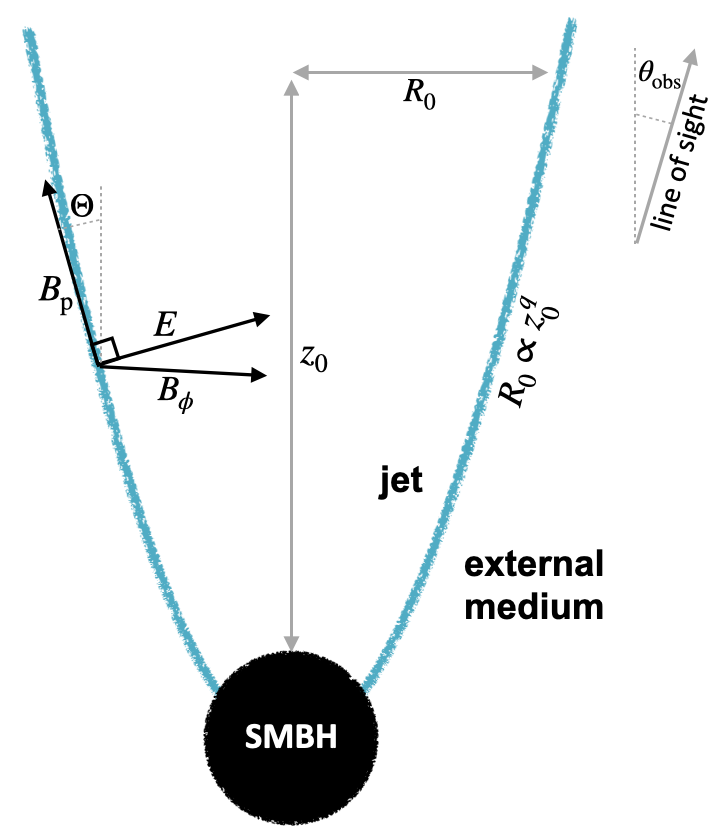}
    \caption{Shape of the jet. The boundary radius, $R_0$ ,is proportional to $z_{0}^q$, where $z_0$ is the distance from the supermassive black hole. The value of the parameter $q<1$ depends on the pressure profile of the external medium that collimates the jet. The line of sight makes an angle $\theta_{\rm obs}$ with respect to the direction of the jet axis.
    }
    \label{fig:scenario}
\end{figure}

According to a widely accepted paradigm, relativistic jets extract the rotational energy of the supermassive black hole via electromagnetic stresses \citep{BlandfordZnajek1977, BlandfordPayne1982, Tchekhovskoy2011}. This process produces magnetized outflows, where most of the energy is carried in the form of Poynting flux. The structure of such Poynting-dominated outflows has been thoroughly investigated both numerically \citep{Komissarov2007, Komissarov2009, Tchekhovskoy2008, Tchekhovskoy09} and analytically \citep{Vlahakis2004, Lyubarsky2009, Lyubarsky2010, Lyubarsky2011}.

\citet{Lyubarsky2009} derived the asymptotic equations that govern axisymmetric stationary outflows collimated by an external medium. These equations are valid far beyond the light cylinder, where the jet is accelerated to large Lorentz factors. The outflow is assumed to be force-free. The properties of the outflow are summarized below.

When the external pressure decreases as a power law, $\mathcal{P}_{\rm ext} \propto z_0^{-\kappa}$, where $z_0$ is the distance from the black hole, the jet transverse radius can be parametrized as $R_{0}\propto z_{0}^{q}$. For $0<\kappa<2$ one has $q=\kappa/4$, whereas for a wind-like medium with $\kappa=2$ one has $1/2<q<1$ (when $\kappa=2$, the value of $q$ depends on the ratio of the magnetic and external pressure near the light cylinder). Very long baseline interferometry imaging of radio emission from AGNs jets suggests $q\sim 0.5-1$ \citep{Mertens2016, Pushkarev2017, Kovalev2020, Boccardi2021}. The shape of the jet is sketched in Fig.~\ref{fig:scenario}.

We adopted a cylindrical coordinate system $(R,\phi,z)$, where $\hat{\mathbf{z}}$ is directed along the jet axis. The electromagnetic fields can be presented as\begin{align}
\label{eq:Efield}
\mathbf{E} & = E_{R} \hat{\mathbf{R}} + E_{z} \hat{\mathbf{z}}\\
\label{eq:Bfield}
\mathbf{B} & = B_{R} \hat{\mathbf{R}} + B_{\phi} \hat{\bm{\phi}} +B_{z} \hat{\mathbf{z}} \;,
\end{align}
where\begin{align}
\label{eq:Ecomp}
E_{R} & = \Omega R B_{\rm p} \cos \Theta\;, & E_{z} & = -\Omega R B_{\rm p} \sin \Theta \\
B_{R} & = B_{\rm p}\sin \Theta\;, & B_{z} & = B_{\rm p} \cos \Theta \;.
\end{align}
The speed of light is set to $c=1$. The angular velocity, $\Omega$, and the poloidal magnetic field, $B_{\rm p}$, are assumed to be independent of $R$. Numerical simulations show that this is a reasonable approximation (see, e.g.,~Fig.~4 of \citealt{Komissarov2007}).

The local jet opening angle, $\Theta,$ is determined by the shape of the magnetic flux surfaces. The boundary radius, $R_0,$ can be parametrized as $\Omega R_{0} = 3^{1 / 4} C ( \Omega z_{0})^{q}$, where $C$ is a constant of order unity \citep{Lyubarsky2009}. At the distance $z_0$ from the black hole, one has $\Theta\simeq| E_{z} / E_{R}|= q R / z_0$, which gives
\begin{equation}
 \Theta =  q \; \frac{R}{R_{0}} \frac{\left(3^{1 / 4} C \right)^{1 / q}}{\left(\Omega R_{0} \right)^{\left(1-q\right) / q}}\;.
\end{equation}
As we discuss in Appendix \ref{sec:Bphi}, in the model of \citet{Lyubarsky2009} the toroidal magnetic field is determined by
\begin{equation}
\label{eq:BminusE}
 \frac{B^{2}_{\phi} - E^{2}}{B^{2}_{\rm p}} = \left( 3^{1 / 4} \; C \right)^{2 / q} \; \frac{q \left(1-q \right)}{3}  \frac{\left(\Omega R\right)^{4}}{\left(\Omega R_{0}\right)^{2 / q}} \;.
\end{equation}

\begin{figure*}
    \centering
    \includegraphics[width=0.49\linewidth]{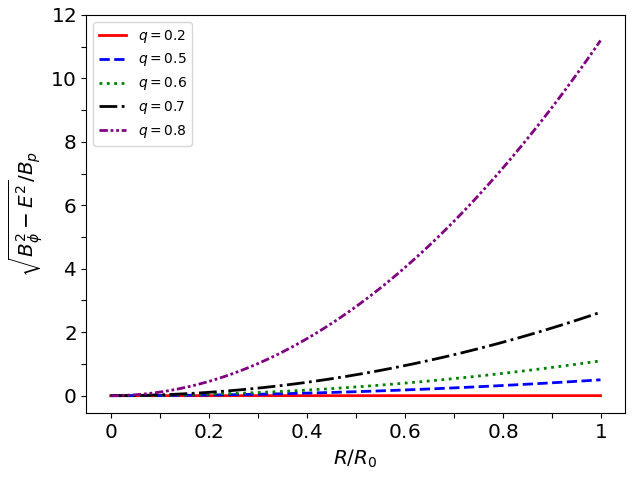}
    \includegraphics[width=0.49\linewidth]{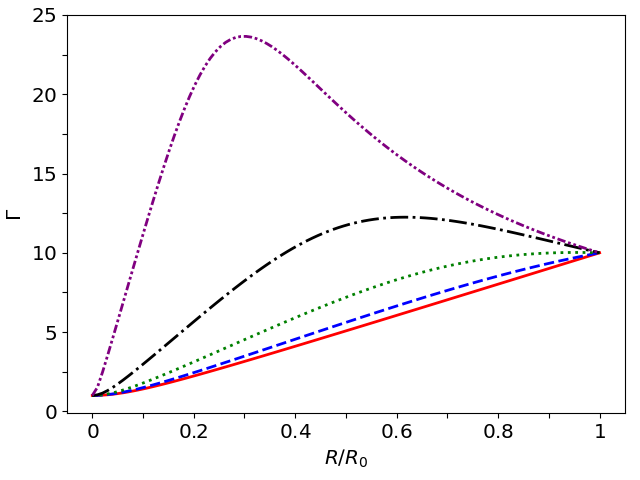}
    \caption{Ratio of the toroidal and poloidal components of the magnetic field in the proper frame of the fluid (left panel) and bulk Lorentz factor of the fluid in the observer's frame (right panel) as a function of the distance, $R,$ from the jet axis.
    }
    \label{fig:gamma}
\end{figure*}

Taking into account that $\Omega R_0\gg 1$ far beyond the light cylinder, from Eqs.~\eqref{eq:Ecomp} and \eqref{eq:BminusE} one can show that $B_\phi\sim E_R\gg B_{\rm p}$ in the observer's frame. However, the poloidal magnetic field cannot be neglected because the toroidal field and the electric field nearly cancel in the proper frame. The left panel of Fig.~\ref{fig:gamma} shows the ratio of the toroidal and poloidal components of the magnetic field in the proper frame as a function of the distance from the jet axis. For small values of $q$ (i.e.,~when the jet is nearly cylindrical) the poloidal component is larger than the toroidal component. For large $q$ (i.e.,~when the jet opening angle increases) the toroidal component starts to dominate, especially near the jet boundary. As we explain in Sect. \ref{sec:pol}, these features are crucial for the polarization of the synchrotron radiation from the jet.

We assumed that the bulk velocity of the fluid, $\mathbf{v}$, is equal to the drift velocity, which is
\begin{equation}
\label{eq:vdrift}
\mathbf{v}= \frac{\mathbf{E}\times\mathbf{B}}{B^{2}} \;.    
\end{equation}
The right panel of Fig.~\ref{fig:gamma} shows the corresponding Lorentz factor, $\Gamma=(1-v^2)^{-1/2}=(1-E^2/B^2)^{-1/2}$. The value of $\Omega R_{0}$ is determined from the condition $\Gamma(R_0)=10$, which is the typical Lorentz factor of AGN jets. For small values of $q$, the Lorentz factor increases linearly with the radius (i.e.,~$\Gamma \sim \Omega R$). For large values of $q$, the Lorentz factor peaks at intermediate radii. One can show that near the jet boundary the Lorentz factor is determined by the curvature of the magnetic flux surfaces. Small and large values of $q$ correspond respectively to the first and second collimation regimes identified by \citet{Lyubarsky2009}.

\section{Polarization}
\label{sec:pol}

\begin{figure*}
    \centering
    \includegraphics[width=0.48\linewidth]{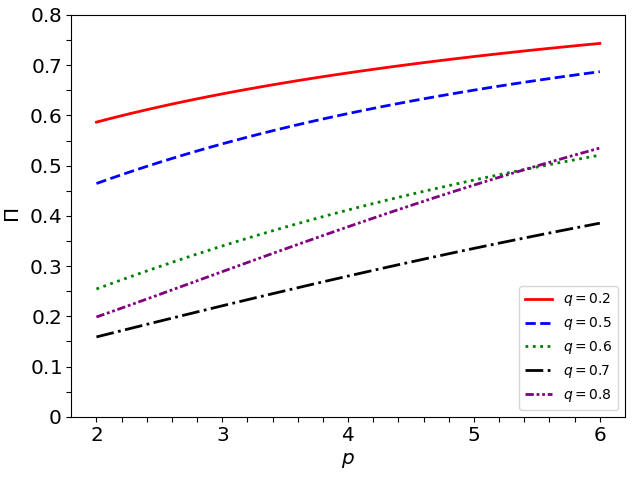}
    \includegraphics[width=0.48\linewidth]{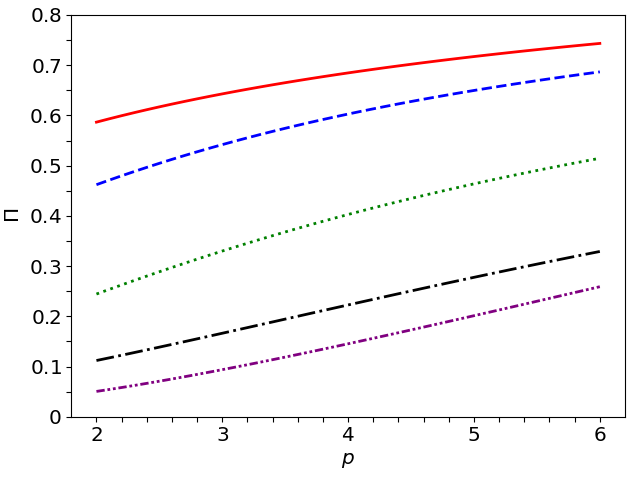}
    \includegraphics[width=0.48\linewidth]{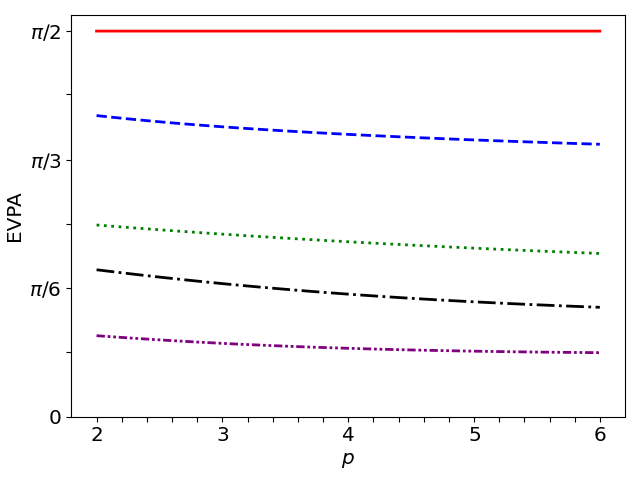}
    \includegraphics[width=0.48\linewidth]{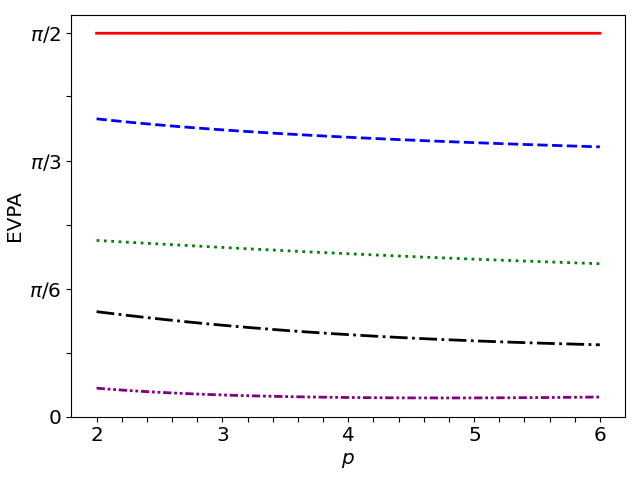}
    \caption{Polarization degree, $\Pi$ (top panels) and EVPA (bottom panels) as a function of the power-law index, $p,$ of the electron energy distribution. When ${\rm EVPA}=0$ and ${\rm EVPA}=\pi/2$, the EVPA is respectively parallel and perpendicular to the projection of the jet axis on the plane of the sky. In the left panels the proper number density is assumed to be constant, whereas in the right panels the magnetization is assumed to be constant. The viewing angle is assumed to be $\theta_{\rm obs}=0.1{\rm\; rad}$.}
    \label{fig:cost}
\end{figure*}

In order to study the polarization of the synchrotron radiation from the jet, we assumed that the energy distribution of the emitting electrons in the proper frame is\begin{equation}
\frac{dN}{d\gamma} = K_{e} \gamma^{-p}\;, 
\end{equation}
where $K_{e}(R,z)$ is the proper electron number density. We also assumed that the distribution is isotropic.

The polarization of the synchrotron radiation from relativistic outflows has been investigated by several authors \citep[e.g.,][]{BlandfordKoenigl79, Pariev2003, Lyutikov2003, Lyutikov05, DelZanna06}. The Stokes parameters per unit length of a stationary jet are given by\footnote{Since we are dealing with ultra-relativistic particles, we neglected circular polarization and set $V=0$.}
\begin{align}
\label{eq:stokes1}
I & = \frac{p + 7/3}{p+1} \; \kappa_{p} \int {\rm d}V \; K_{e} \mathcal{D}^{\left( 3 + p\right)/2} \;  \left| \mathbf{B}' \times \hat{\mathbf{n}}' \right|^{\left( p+1 \right)/2} \\
\label{eq:stokes2}
Q & = \kappa_{p} \int {\rm d}V \; K_{e} \mathcal{D}^{\left( 3 + p\right)/2} \;  \left| \mathbf{B}' \times \hat{\mathbf{n}}' \right|^{\left( p+1 \right)/2} \; \cos 2 \chi \\
\label{eq:stokes3}
U & = \kappa_{p} \int {\rm d}V \; K_{e} \mathcal{D}^{\left( 3 + p\right)/2} \;  \left| \mathbf{B}' \times \hat{\mathbf{n}}' \right|^{\left( p+1 \right)/2}  \sin 2 \chi \; ,
\end{align}
where $\mathcal{D}$ is the Doppler factor, $|\mathbf{B}' \times \hat{\mathbf{n}}'|$ is the strength of the magnetic field component perpendicular to the line of sight, and $\chi$ is the angle between the polarization vector from a volume element and some reference direction in the plane of the sky that will be specified below. Primed quantities refer to the proper frame of the fluid. The integration volume is\footnote{The Stokes parameters per unit jet length provide good estimates of $\Pi$ and EVPA when the emission region is not very extended. For example, taking ${\rm d}V = R\; {\rm d}R \; {\rm d}\phi\; {\rm d}z$, where $z_0-\Delta z<z<z_0$, we find that $\Pi$ and EVPA change by $<20\%$ for $\Delta z/z_0<0.5$. Our results do not depend on the exact location of the emission region (i.e.,~fixing $\Gamma (R_0)=10$ one finds that $\Pi$ and EVPA are independent of $z_0$).} ${\rm d}V = R\; {\rm d}R \; {\rm d}\phi$, where $0<\phi<2\pi$ and $0<R<R_{0}$. We also defined
\begin{align}
\kappa_{p} & = \frac{C_{p} \; \nu ^{- \left( p-1 \right)/2} }{D^{2}\left( 1 + z \right)^{\left( 3 + p\right)/2}}\\
C_{p} & = \frac{\sqrt{3}}{4} \Gamma_{\rm E} \left(  \frac{3p-1}{12}\right)  \Gamma_{\rm E} \left(  \frac{3p+7}{12}\right) \frac{e^{3}}{m_{e}} \left( \frac{3e}{2 \pi m^{3}_{e} }\right)^{\left(p-1\right)/2},
\end{align}
where $e$ and $m_{e}$ are the charge and mass of the electron, $\nu$ is the observed radiation frequency, $z$ is the cosmological redshift, $D$ is the luminosity distance to the source, and $\Gamma_{\rm E}$ is the Euler gamma function. The degree of linear polarization is
\begin{equation}
\label{PI}
\Pi = \frac{\sqrt{Q^{2} + U^{2}}}{I}\;, 
\end{equation}
and the EVPA, $\Psi,$ of the synchrotron radiation from the entire emission region (assumed to be unresolved) is given by
\begin{equation}
\cos 2 \Psi = \frac{Q}{\sqrt{Q^{2} + U^{2}}} \;, \qquad  \sin 2 \Psi = \frac{U}{\sqrt{Q^{2} + U^{2}}} \;, 
\end{equation}
where $0 < \Psi < \pi$.

Now we express the quantities that appear in Eqs.~\eqref{eq:stokes1}-\eqref{eq:stokes3} as a function of the electromagnetic fields in the observer's frame (our final expressions are given by Eqs.~\ref{eqn:chi2}-\ref{eq:Bperp} below). We closely followed the framework developed by \citet{DelZanna06}. The viewing angle $\theta_{\rm obs}$ is measured with respect to the direction of the jet axis, $\hat{\mathbf{z}}$. Then, the unit vector directed toward the observer is $\hat{\mathbf{n}}= \sin{\theta_{\rm obs}}\cos{\phi}\; \hat{\mathbf{R}} - \sin{\theta_{\rm obs}} \sin{\phi}\; \hat{\bm{\phi}}+ \cos{\theta_{\rm obs}}\hat{\mathbf{z}}$. The polarization vector of the radiation from a volume element is \citep{Lyutikov2003, DelZanna06}
\begin{equation}
\label{eqn:polvec}
\hat{\mathbf{e}} = \frac{\hat{\mathbf{n}} \times \mathbf{q}}{\sqrt{  q^{2} - \left( \mathbf{q} \cdot \hat{\mathbf{n}} \right)^{2} }} \;,  \qquad \mathbf{q} = \mathbf{B} - \hat{\mathbf{n}} \times \mathbf{E} \;.
\end{equation}
We took as a reference direction the projection of the jet axis on the plane of the sky, which is $\hat{\mathbf{l}}=[(  \hat{\mathbf{z}} \cdot \hat{\mathbf{n}})\hat{\mathbf{n}}-\hat{\mathbf{z}}]/\sqrt{1-(\hat{\mathbf{z}} \cdot \hat{\mathbf{n}})^2}$ (note that $\hat{\mathbf{l}}$ is the unit vector orthogonal to $\hat{\mathbf{n}}$, and coplanar to $\hat{\mathbf{n}}$ and $\hat{\mathbf{z}}$). Then, one has $\cos\chi=\hat{\mathbf{e}} \cdot \hat{\mathbf{l}}$, and $\sin\chi=\hat{\mathbf{e}} \cdot (\hat{\mathbf{l}} \times \hat{\mathbf{n}})$, or equivalently
\begin{equation}
\label{eqn:chi}
\cos\chi = \frac{\mathbf{q}\cdot \left(\hat{\mathbf{l}} \times\hat{\mathbf{n}}\right)}{\sqrt{  q^{2} - \left( \mathbf{q} \cdot \hat{\mathbf{n}} \right)^{2} }}\;, \qquad
\sin\chi = -\frac{\mathbf{q}\cdot\hat{\mathbf{l}}}{\sqrt{  q^{2} - \left( \mathbf{q} \cdot \hat{\mathbf{n}} \right)^{2} }} \;.
\end{equation}
Since $\hat{\mathbf{l}}\times\hat{\mathbf{n}}$, $\hat{\mathbf{l}}$, $\hat{\mathbf{n}}$ are mutually orthogonal, any vector $\mathbf{X}$ can be conveniently decomposed as $\mathbf{X}= X_1\hat{\mathbf{l}} \times\hat{\mathbf{n}}+ X_2\hat{\mathbf{l}}+X_3\hat{\mathbf{n}}$, where $X_1=\mathbf{X}\cdot (\hat{\mathbf{l}} \times\hat{\mathbf{n}})$, $X_2=\mathbf{X}\cdot\hat{\mathbf{l}}$, $X_3=\mathbf{X}\cdot \hat{\mathbf{n}}$. Below we need the first and second components of $\mathbf{q}=\mathbf{B}-\hat{\mathbf{n}}\times\mathbf{E}$, which are
\begin{align}
q_1 & =  \left( E_{R} \cos \theta_{\rm obs} - B_{\phi} \right)\cos \phi - E_{z} \sin \theta_{\rm obs} - B_{R} \sin \phi \\
q_2 & = \left( E_{R} - B_{\phi} \cos \theta_{\rm obs} \right) \sin \phi - B_{z} \sin \theta_{\rm obs} + B_{R} \cos \theta_{\rm obs} \cos \phi \;,
\end{align}
and the third component of $\mathbf{v}=\mathbf{E} \times \mathbf{B}/B^2$, which is
\begin{equation}
v_3 = \frac{\left[\left(B_z E_R-B_R E_z\right)\sin\phi-B_\phi E_z\cos\phi\right]\sin\theta_{\rm obs} + B_\phi E_R\cos\theta_{\rm obs}}{B^2}
.\end{equation}
Equation~\eqref{eqn:chi} gives $\cos\chi=q_1/\sqrt{q_1^2+q_2^2}$ and $\sin\chi=-q_2/\sqrt{q_1^2+q_2^2}$. Then one has
\begin{equation}
\label{eqn:chi2}
\cos 2\chi = \frac{q_1^2-q_2^2}{q_1^2+q_2^2}\;, \qquad
\sin 2\chi = -\frac{2q_1q_2}{q_1^2+q_2^2} \;.
\end{equation}
The Doppler factor, $\mathcal{D} =[\Gamma ( 1 - \mathbf{v} \cdot \hat{\mathbf{n}})]^{-1}$, can be presented as
\begin{equation} 
\label{eq:doppler1}
\mathcal{D} = \frac{1}{\Gamma\left(1-v_3\right)} \;.
\end{equation}
Taking into account that $\mathbf{v} \cdot \mathbf{B}=0$, the strength of the magnetic field component perpendicular to the line of sight is \citep{DelZanna06}: $| \mathbf{B}' \times \hat{\mathbf{n}}'|=\Gamma^{-1}\sqrt{B^2-\mathcal{D}^2(\mathbf{B} \cdot\hat{\mathbf{n}})^2}$. As we demonstrate in Appendix \ref{sec:identities}, this expression is equivalent to
\begin{equation}
\label{eq:Bperp}
\left| \mathbf{B}' \times \hat{\mathbf{n}}' \right| = \mathcal{D} \sqrt{q_1^2+q_2^2} \;.    
\end{equation}

To calculate the Stokes parameters, one needs to specify the electron number density $K_e$. We considered two different scenarios: a constant number density, $K_e=K_0$ (Sect. \ref{sec:uni}) and a constant magnetization, $K_e\propto B'^2$, where $B'=B/\Gamma$ is the magnetic field strength in the proper frame (Sect. \ref{sec:mag}). We assumed that the viewing angle is $\theta_{\rm obs} = 1/\Gamma(R_0)=0.1{\rm\; rad}$. The effect of changing the viewing angle is discussed in Appendix \ref{sec:thetaobs}.

\subsection{Constant number density}
\label{sec:uni}

In the left panels of Fig.~\ref{fig:cost} we show the polarization degree $\Pi$ and the EVPA obtained for a constant proper number density (i.e.,~$K_{e}=K_0$) as a function of the power-law index, $p,$ of the electron energy distribution. The spectrum of the emitted radiation is $F_\nu\propto \nu^{-\alpha}$, where $\alpha=(p-1)/2$. The typical value of the optical spectral index of HSP blazars is $\alpha\sim 0.5$ \citep{fossati98, Abdo2011, ghisellini17}, which is obtained for $p=2$. The X-ray spectral index is typically within the range $\alpha\sim 1.5-2.5$, which is obtained for $p=4-6$. Hereafter, we assume the optical and X-ray polarization to be respectively $\Pi_{\rm O}=\Pi_{p=2}$ and $\Pi_{p=4}<\Pi_{\rm X}<\Pi_{p=6}$.

For small values of $q$, the polarization degree is close to the maximal one, which is $\Pi_{\rm max}=(p+1)/(p+7/3)$. For example, considering a jet with $q=0.2$ we find\footnote{The X-ray polarization degree depends on the power-law index of the electron energy distribution. $\Pi_{\rm X}=0.68$ and $\Pi_{\rm X}=0.74$ are obtained for $p=4$ and $p=6,$ respectively.} $\Pi_{\rm X}=0.68-0.74$ and $\Pi_{\rm O}=0.58$, whereas $\Pi_{\rm max,X}= 0.79-0.84$ and $\Pi_{\rm max,O}= 0.69$. The polarization degree is weakly chromatic, as $\Pi_{\rm X}/\Pi_{\rm O}=1.2$. Considering $q=0.05-0.1$, the polarization degree changes by less than a few percent with respect to the model with $q=0.2$. The EVPA is nearly perpendicular to the projection of the jet axis on the plane of the sky, as expected since in the proper frame the magnetic field is predominantly poloidal (see Fig.~\ref{fig:gamma}).

For large values of $q$, the polarization degree is much smaller than $\Pi_{\rm max}$. For $q=0.6$ we find $\Pi_{\rm X}=0.41-0.52$ and $\Pi_{\rm O}=0.25$, and for $q=0.8$ we find $\Pi_{\rm X}= 0.38-0.54$ and $\Pi_{\rm O}= 0.20$. The polarization degree is strongly chromatic, as $\Pi_{\rm X}/\Pi_{\rm O}=1.6-2.7$. This effect is entirely due to the softening of the electron energy distribution at high energies (we assumed $p=2$ for the optical emitting electrons, and $p=4-6$ for the X-ray emitting electrons). The chromaticity is much stronger than for a uniform magnetic field, where the polarization degree is $\Pi_{\rm max}=(p+1)/(p+7/3)$ (Fig.~5 of \citealt{Lyutikov2003} shows that the polarization degree can increase with $p$ more rapidly than $\Pi_{\rm max}$, consistent with our results). The EVPA is nearly parallel to the projection of the jet axis on the plane of the sky, as expected since in the proper frame the magnetic field is predominantly toroidal. For $q\gtrsim 0.8$ the polarization degree slightly increases, whereas the EVPA shows a continuous transition to ${\rm EVPA}=0$. As we discuss in Appendix \ref{sec:special}, an abrupt transition from ${\rm EVPA}=\pi/2$ to ${\rm EVPA}=0$ is possible only for some special jet shapes.

\subsection{Constant magnetization}
\label{sec:mag}

In the right panels of Fig.~\ref{fig:cost} we show the polarization degree $\Pi$ and the EVPA obtained for a constant magnetization (i.e.,~$K_{e}\propto B'^2=B^2/\Gamma^2$). For small values of $q$, one has $B_\phi^2-E^2\ll B_{\rm p}^2$ (see Fig.~\ref{fig:gamma}), and therefore $B'^2=B^2-E^2\sim B_{\rm p}^2$. Since $B_{\rm p}$ is independent of $R$, the polarization degree and the EVPA are similar to the scenario with constant number density.

For large values of $q$, one has $B_\phi^2-E^2\gg B_{\rm p}^2$, and therefore $B'^2\sim B_\phi^2-E^2$. Since $B_\phi^2-E^2\propto R^4$ (see Eq.~\ref{eq:BminusE}), the proper number density has a sharp peak near the jet boundary. The polarization degree is smaller with respect to the scenario with a constant number density. For $q=0.6$ we find $\Pi_{\rm X}=0.40-0.51$ and $\Pi_{\rm O}=0.24$, and for $q=0.8$ we find $\Pi_{\rm X}=0.15-0.26$ and $\Pi_{\rm O}=0.05$. Larger values of $q$ would give a lower polarization degree. For example, for $q=0.9$ we find $\Pi_{\rm X}=0.067-0.15$ and $\Pi_{\rm O}=0.016$. In this case, the chromaticity is very strong, as $\Pi_{\rm X}/\Pi_{\rm O}=4.2-9.3$. However, the peak bulk Lorentz factor would be very large.

\section{Discussion}
\label{sec:disc}

We studied the polarization of the synchrotron radiation from HSP blazars. We used the axisymmetric stationary model of \citet{Lyubarsky2009} to calculate the jet electromagnetic fields, as appropriate for blazars in a quiescent state. In this model, the transverse size of the jet can be parameterized as $R_0\propto z_0^q$, where $z_0$ is the distance from the black hole. We considered two simple scenarios to populate the jet with nonthermal electrons (we assumed that either the magnetization or the electron proper number density across the jet is constant). The shape of the jet has a dramatic impact on the polarization:
\begin{itemize}
\item For $q\sim 0.2$ (a nearly cylindrical jet), the polarization degree is large and weakly chromatic (the X-ray and optical polarization degrees are respectively $\Pi_{\rm X}\sim 70\%$ and $\Pi_{\rm O}\sim 60\%$). The EVPA is nearly perpendicular to the projection of the jet axis on the plane of the sky.
\item For $q\sim 0.6-0.8$ (a nearly parabolic jet), the polarization degree is lower and becomes strongly chromatic. We find $\Pi_{\rm X}\sim 15-50\%$ and $\Pi_{\rm O}\sim 5-25\%$ (the ratio is typically $\Pi_{\rm X}/\Pi_{\rm O}\sim 2-5$; however, the ratio can reach $\Pi_{\rm X}/\Pi_{\rm O}\sim 9$ for $q=0.9$). The EVPA is nearly parallel to the projection of the jet axis on the plane of the sky.
\end{itemize}
We emphasize that in our model the optical and X-ray emitting electrons are co-spatial and only differ in terms of the power-law index of their energy distributions. The chromaticity is due to the softening of the electron distribution at high energies and is much stronger than for a uniform magnetic field. The polarization degree and the EVPA are determined by the global structure of the electromagnetic fields within the jet (the effect of a random component of the fields is discussed below).

Our results are important for interpreting multifrequency polarimetric observations of HSP blazars \citep{liodakis22, digesu22, ehlert2023, Errando2024, Kim2024}. The observed X-ray polarization degree is $\Pi_{\rm X}\sim 10-20\%$, whereas the optical polarization degree is significantly smaller ($\Pi_{\rm X}/\Pi_{\rm O}\sim 2-7$). The EVPA seems to align with the projection of the jet axis on the plane of the sky.\footnote{The situation is far from clear, especially for the blazars 1ES 0229+200 \citep{ehlert2023} and Mrk 421 \citep{digesu22}. As noted in \cite{digesu22}, the jet could bend by tens of degrees between the inner region (where the X-ray and optical emission are likely produced) and the outer region imaged in the radio band.} Our results show that the jet should be nearly parabolic ($q\sim 0.6-0.8$), consistent with very long baseline interferometry observations of AGN jets on larger spatial scales \citep{Mertens2016, Pushkarev2017, Kovalev2020, Boccardi2021}.

Current multifrequency polarimetric observations of HSP blazars can hardly constrain the particle acceleration process. In our model the jet is Poynting-dominated, and therefore nonthermal particle acceleration is likely powered by magnetic reconnection rather than shocks \citep{Sironi15b}. The most energetic electrons trace the distribution of the reconnecting current sheets, as their cooling time is short. The Kelvin-Helmholtz instability at the interface between the jet and the external medium can produce current sheets near the jet boundary \citep{Sironi2021}. Instead, the kink instability can produce a more uniform distribution of current sheets across the jet \citep{Alves2018, Davelaar2020}. In the scenarios that we explored, the polarization degree and the EVPA are not very sensitive to the radial distribution of the electrons. As such, it is difficult to say whether reconnection is triggered by the Kelvin-Helmholtz or the kink instability.

We conclude by discussing some directions for future research:

 (i) The electromagnetic fields may have a random component that changes on small spatial scales. If the random component is isotropic in the proper frame, and its strength is comparable to the regular component, the polarization degree decreases by a factor of $\sim 2$, whereas the EVPA does not change \citep{Granot2003, Bandiera2016}. A time-dependent random magnetic field can produce fluctuations of the polarization degree and EVPA \citep{marscher14}.

(ii) Nonthermal electrons could be continuously accelerated at a certain distance from the black hole and then advected along the jet for a cooling length. Since the cooling length depends on the electron energy, this leads to ``energy stratification,'' as previously discussed for the shock acceleration scenario \citep{tavecchio18, tavecchio20}. We emphasize that some energy stratification is expected in any conceivable scenario, as particles with a longer cooling time are advected farther away along the jet. The polarization degree is expected to decrease as the emission region becomes more spatially extended. This effect is chromatic ($\Pi_{\rm O}$ decreases more than $\Pi_{\rm X}$ because the corresponding electron cooling time is longer by a factor of $\sqrt{\nu_{\rm X}/\nu_{\rm O}}\sim 100$).

(iii) Both optical and X-ray EVPA swings have now been observed \citep{Marscher08, marscher10, Abdo2010b, Larionov2013, blinov15, blinov16, blinov18, Kiehlmann2016, digesu23, middei23a}. To explain the EVPA swings, one should abandon the axisymmetric stationary jet model that we used throughout this paper. The fact that optical and X-ray EVPA swings are not simultaneous suggests that, due to energy stratification, the emitting electrons are not exactly co-spatial \citep{digesu23, middei23a}.

\begin{acknowledgements}
We are grateful to the anonymous referee for their constructive comments. We acknowledge insightful discussions with Rino Bandiera, Niccolò Bucciantini, Luca Del Zanna, and Jonathan Granot. We acknowledge financial support from the Marie Sk{\l}odowska-Curie Grant 101061217 (PI E.~Sobacchi) and from a INAF Theory Grant 2022 (PI F.~Tavecchio). This work has been funded by the European Union-Next Generation EU, PRIN 2022 RFF M4C21.1 (2022C9TNNX).
\end{acknowledgements}

\bibliographystyle{aa}
\bibliography{tavecchio}

\appendix

\section{Effect of the viewing angle}
\label{sec:thetaobs}

\begin{figure}
    \centering
    \includegraphics[width=\linewidth]{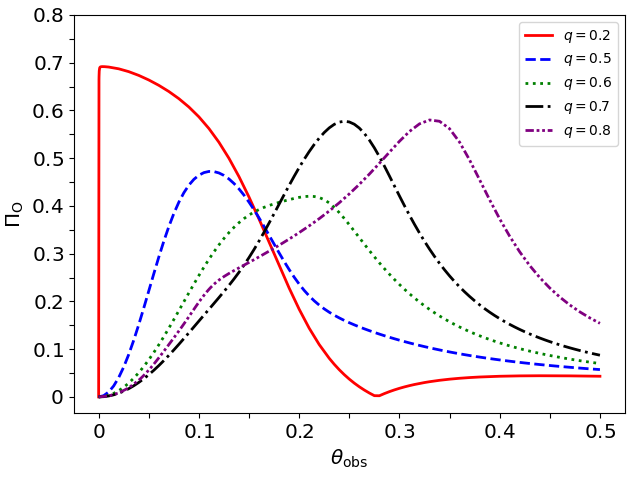}
    \includegraphics[width=\linewidth]{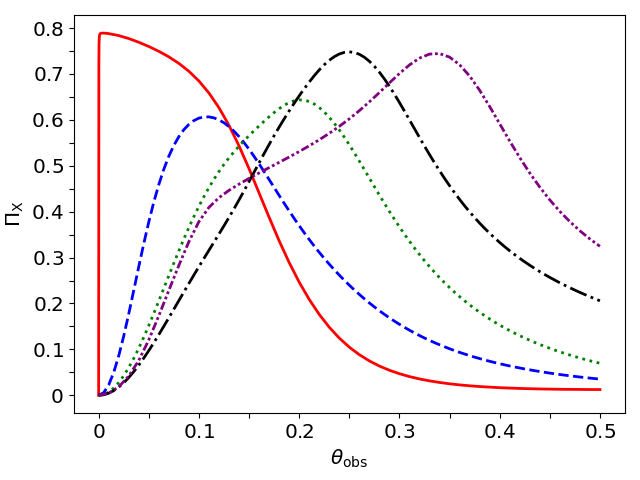}
    \caption{Optical polarization degree, $\Pi_{\rm O}=\Pi_{p=2}$ (top panel) and X-ray polarization degree, $\Pi_{\rm X}=\Pi_{p=4}$ (bottom panel) as a function of the viewing angle, $\theta_{\rm obs}$, which is expressed in radians. The proper number density of the emitting electrons is assumed to be constant.
    }
    \label{fig:theta_obs}
\end{figure}

We investigated the dependence of the polarization degree on the viewing angle. We assumed that the proper number density of the emitting electrons is constant. In the top panel of Fig.~\ref{fig:theta_obs} we show the optical polarization degree, and in the bottom panel we show the X-ray polarization degree. Both $\Pi_{\rm O}$ and $\Pi_{\rm X}$ show a similar dependence on $\theta_{\rm obs}$. The most important difference is that $\Pi_{\rm O}$ is smaller than $\Pi_{\rm X}$ for a given $\theta_{\rm obs}$.

The polarization degree vanishes for $\theta_{\rm obs}=0$, as expected from symmetry considerations. For $q\sim 0.2$, the polarization degree rises almost abruptly to $\Pi_{\rm max}$, and then decreases for large $\theta_{\rm obs}$. For $q\sim 0.5-0.8$, the rise is more gradual, and the peak polarization is smaller than $\Pi_{\rm max}$.

\section{Special jet shapes}
\label{sec:special}

In the general case discussed in Sect. \ref{sec:pol}, the Stokes parameter $U$ does not vanish. However, for the special jet shapes discussed below one finds $U=0$. When $U=0$, the EVPA can be either parallel or perpendicular to the projection of the jet axis on the plane of the sky.

\citet{Lyutikov2003} considered a conical jet with $B_R=B_z=0$. In this case, one has
\begin{align}
\label{eq:q1con}
q_1 & = \left( E_{R} \cos \theta_{\rm obs} - B_{\phi} \right)\cos \phi - E_{z} \sin \theta_{\rm obs} \\
q_2 & = \left( E_{R} - B_{\phi} \cos \theta_{\rm obs} \right) \sin \phi \\
v_3 & = \frac{B_\phi E_R\cos\theta_{\rm obs}- B_\phi E_z\cos\phi\sin\theta_{\rm obs}}{B^2} \;.
\end{align}
For a given viewing angle $\theta_{\rm obs}$, the Doppler factor is a function of $R$ and $\cos\phi$:
\begin{equation}
\label{eq:Dcon}
\mathcal{D} = \frac{1}{\Gamma\left(1-v_3\right)} = f_1\left(R,\cos\phi\right) \;.
\end{equation}
Since $q_1^2+q_2^2$ is also a function of $R$ and $\cos\phi$, one has
\begin{equation}
\left|\mathbf{B}'\times\hat{\mathbf{n}}'\right| = \mathcal{D}\sqrt{q_1^2+q_2^2} = f_2\left(R, \cos\phi\right)\;.
\end{equation}
Finally, one finds
\begin{equation}
\label{eq:sincon}
\sin 2\chi = -\frac{2q_1q_2}{q_1^2+q_2^2} = f_3\left(R,\cos\phi\right) \sin\phi\;.
\end{equation}
The functions $f_1$, $f_2$, $f_3$ are defined through Eqs.~\eqref{eq:Dcon}-\eqref{eq:sincon}. Substituting Eqs.~\eqref{eq:Dcon}-\eqref{eq:sincon} into Eq.~\eqref{eq:stokes3}, one finds
\begin{equation}
\label{eq:Uconical}
U=\int_0^{R_0}R{\rm d}R\int_0^{2\pi}{\rm d}\phi \;F(R,\cos\phi) \sin\phi\;,
\end{equation}
where the function $F$ can be expressed as a combination of $f_1$, $f_2$, $f_3$. It is clear that $U=0$ for any function $F$.

\citet{Pariev2003} and \citet{Lyutikov05} considered a cylindrical jet with $E_{z}=B_{R}=0$. In this case, one has
\begin{align}
q_1 & = \left( E_{R} \cos \theta_{\rm obs} - B_{\phi} \right)\cos \phi \\
q_2 & = \left( E_{R} - B_{\phi} \cos \theta_{\rm obs} \right) \sin \phi - B_{z} \sin \theta_{\rm obs} \\
v_3 & = \frac{B_\phi E_R\cos\theta_{\rm obs} + B_z E_R\sin\phi\sin\theta_{\rm obs}}{B^2}\;.
\end{align}
The Doppler factor is a function of $R$ and $\sin\phi$:
\begin{equation}
\label{eq:Dcyl}
\mathcal{D} = \frac{1}{\Gamma\left(1-v_3\right)} = g_1\left(R,\sin\phi\right) \;.
\end{equation}
Since $q_1^2+q_2^2$ is also a function of $R$ and $\sin\phi$, one has
\begin{equation}
\left|\mathbf{B}'\times\hat{\mathbf{n}}'\right| = \mathcal{D}\sqrt{q_1^2+q_2^2} = g_2\left(R, \sin\phi\right)\;.
\end{equation}
Finally, one finds\begin{equation}
\label{eq:sincyl}
\sin 2\chi = -\frac{2q_1q_2}{q_1^2+q_2^2} = g_3\left(R,\sin\phi\right) \cos\phi\;.
\end{equation}
Substituting Eqs.~\eqref{eq:Dcyl}-\eqref{eq:sincyl} into Eq.~\eqref{eq:stokes3}, one finds\begin{equation}
U=\int_0^{R_0}R{\rm d}R\int_0^{2\pi}{\rm d}\phi \;G(R,\sin\phi) \cos\phi\;,
\end{equation}
which gives $U=0$.

\section{Details of the jet structure}
\label{sec:Bphi}

When the jet is accelerated to bulk Lorentz factors $\Gamma\gg 1$, Eq.~(29) of \citet{Lyubarsky2009} gives
\begin{equation}
\label{eq:D1}
\frac{B_\phi^2-E^2}{B_{\rm p}^2} = \frac{\Omega^2R^2}{\Gamma^2} -1 \;. 
\end{equation}
Following \citet{Lyubarsky2009}, we defined $X=\Omega R$ and $Z=\Omega z_0$. The shape of the flux surfaces was determined by their Eq.~(74), which gives $\psi/\psi_0 = X^2/\sqrt{3}Y^2$, where $Y=CZ^q$. Substituting this expression into their Eq.~(75), one finds\begin{equation}
\frac{\Omega^2R^2}{\Gamma^2} -1 = \frac{q\left(1-q\right)}{3} \frac{X^4}{Z^2} \;.
\end{equation}
Since the background magnetic field is assumed to be uniform, one has $\psi/\psi_0=R^2/R_0^2=X^2/\Omega^2R_0^2$. As discussed above, one has also $\psi/\psi_0= X^2/\sqrt{3}C^2Z^{2q}$. Then, one finds
\begin{equation}
\label{eq:D3}
Z^2 = \left(\frac{\Omega R_0}{3^{1/4}C}\right)^{2/q} \;.
\end{equation}
Eq.~\eqref{eq:BminusE} immediately follows from Eqs.~\eqref{eq:D1}-\eqref{eq:D3}.

\section{Perpendicular magnetic field}
\label{sec:identities}

We calculated the strength of the magnetic field component perpendicular to the line of sight. First of all, we demonstrated two useful identities. The first identity is
\begin{equation}
\label{eq:C1}
\left[\left(\mathbf{E}\times\mathbf{B}\right)\cdot\hat{\mathbf{n}}\right]^2 = E^2B^2 - \left(\mathbf{B}\cdot\hat{\mathbf{n}} \right)^2 E^2 - \left(\mathbf{E}\cdot\hat{\mathbf{n}}\right)^2 B^2 \;,
\end{equation}
which can be demonstrated as follows. Consider the vector $\mathbf{P}=\left(\mathbf{E}\times\mathbf{B}\right)\times \hat{\mathbf{n}}$. One has
\begin{equation}
\label{eq:C2}
P^2 = \left(\mathbf{E}\times\mathbf{B}\right)^2 - \left[\left(\mathbf{E}\times\mathbf{B}\right)\cdot\hat{\mathbf{n}}\right]^2 = E^2B^2 - \left[\left(\mathbf{E}\times\mathbf{B}\right)\cdot\hat{\mathbf{n}}\right]^2 \;,
\end{equation}
where we used the fact that $\mathbf{E}\cdot\mathbf{B}=0$. Since $\mathbf{P}=(\mathbf{E}\cdot\hat{\mathbf{n}})\mathbf{B}-(\mathbf{B}\cdot\hat{\mathbf{n}})\mathbf{E}$, one also has
\begin{equation}
\label{eq:C3}
P^2 = (\mathbf{B}\cdot\hat{\mathbf{n}})^2E^2 + (\mathbf{E}\cdot\hat{\mathbf{n}})^2B^2 \;.
\end{equation}
Equation \eqref{eq:C1} immediately follows from Eqs.~\eqref{eq:C2}-\eqref{eq:C3}. The second identity, which is obtained using the first one, is
\begin{align}
\notag
\frac{1}{\Gamma^2\mathcal{D}^2} & = \left[1-\frac{\left(\mathbf{E}\times\mathbf{B}\right)\cdot\hat{\mathbf{n}}}{B^2}\right]^2 = \\
\label{eq:C4}
& = 1-\frac{2 \left( \mathbf{E} \times\mathbf{B} \right)\cdot\hat{\mathbf{n}}}{B^2} + \frac{E^2}{B^2} -\frac{\left(\mathbf{B}\cdot\hat{\mathbf{n}} \right)^2 E^2}{B^4} -\frac{\left(\mathbf{E}\cdot\hat{\mathbf{n}}\right)^2}{B^2}.
\end{align}

The strength of the magnetic field component perpendicular to the line of sight is $| \mathbf{B}' \times \hat{\mathbf{n}}'|= \Gamma^{-1} \sqrt{B^2-\mathcal{D}^2(\mathbf{B} \cdot\hat{\mathbf{n}})^2}$. This equation can be presented as
\begin{align}
\notag
\left| \mathbf{B}' \times \hat{\mathbf{n}}'\right| & = \mathcal{D} \sqrt{\frac{B^2}{\Gamma^2\mathcal{D}^2} - \frac{\left(\mathbf{B} \cdot\hat{\mathbf{n}}\right)^2}{\Gamma^2}} = \\
\notag
& = \mathcal{D} \sqrt{\frac{B^2}{\Gamma^2\mathcal{D}^2} - \left(\mathbf{B} \cdot\hat{\mathbf{n}}\right)^2+ \frac{\left(\mathbf{B} \cdot\hat{\mathbf{n}}\right)^2E^2}{B^2}} =\\
\label{eq:C5}
& = \mathcal{D}\sqrt{B^2 -\left(\mathbf{B}\cdot\hat{\mathbf{n}} \right)^2 + E^2 -\left(\mathbf{E}\cdot\hat{\mathbf{n}}\right)^2 -2\left( \mathbf{E} \times\mathbf{B} \right)\cdot\hat{\mathbf{n}}  } \;,
\end{align}
where we used Eq.~\eqref{eq:C4} to obtain the final expression. Now consider the vector $\mathbf{q}=\mathbf{B}-\hat{\mathbf{n}}\times\mathbf{E}$. One has
\begin{align}
\label{eq:C6}
q^2 & = B^2+E^2 -\left(\mathbf{E}\cdot\hat{\mathbf{n}}\right)^2 -2\left( \mathbf{E} \times\mathbf{B} \right)\cdot\hat{\mathbf{n}} \\
\label{eq:C7}
\left(\mathbf{q}\cdot\hat{\mathbf{n}}\right)^2 & = \left( \mathbf{B} \cdot\hat{\mathbf{n}}\right)^2 \;.
\end{align}
From Eqs.~\eqref{eq:C5}-\eqref{eq:C7}, it follows that
\begin{equation}
\left| \mathbf{B}' \times \hat{\mathbf{n}}'\right| = \mathcal{D}\sqrt{q^2-\left(\mathbf{q}\cdot\hat{\mathbf{n}}\right)^2} \;,
\end{equation}
which is equivalent to Eq.~\eqref{eq:Bperp}.

\end{document}